\RequirePackage{fix-cm}
\RequirePackage{amsmath}
\documentclass[smallextended]{svjour3}       
\smartqed  
\usepackage{graphicx}
\usepackage{amssymb}
\usepackage{epstopdf}
\usepackage{color}
\usepackage[font=small]{subcaption}
\begin{document}

\title{Microscopic study of static and dynamical properties of dilute one-dimensional soft bosons 
}

\titlerunning{One-dimensional soft bosons}        

\author{M.~Teruzzi         \and
        D.E.~Galli        \and
        G.~Bertaina
}

\authorrunning{M. Teruzzi et al.} 

\institute{M. Teruzzi \at 
           International School for Advanced Studies (SISSA), Via Bonomea 265, Trieste, Italy\\ 
           \and
           D.E. Galli and G. Bertaina \at
           Dipartimento di Fisica, Universit\`a degli Studi di Milano, via Celoria 16, I-20133 Milano, Italy\\
           \email{gianluca.bertaina@unimi.it}  
}              

\date{Received: date / Accepted: date}

\maketitle

\begin{abstract}
We study static properties and the dynamical structure factor of zero-temperature dilute bosons interacting via a soft-shoulder potential in one dimension. Our approach is fully microscopic and employs state-of-the-art quantum Monte Carlo and analytic continuation techniques. By increasing the interaction strength, our model reproduces the Lieb-Liniger gas, the Tonks-Girardeau and the Hard-Rods models.
\end{abstract}

\section{Introduction} \label{sez:introduction}
Thanks to the progress achieved in the manipulation of ultracold gases with magneto-optical traps, quasi-one-dimensional systems \cite{cazalilla_one_2011} and systems of Rydberg atoms with a hard-core interaction \cite{schaus_observation_2012} have been experimentally realized. Rydberg atoms are highly excited electronic states of alkali atoms, with a large size of the electronic cloud. In addition, recent theoretical \cite{henkel_three-dimensional_2010,Cinti,SaccaniMoroniBononsegnisoft} and experimental \cite{zeiher_many-body_2016} efforts have been put in the study of repulsive finite-range interactions with ensembles of \textit{dressed} Rydberg atoms, which are superpositions of the ground state and the above mentioned excited state, coupled via a two-photon Rabi process of frequency $\Omega$ and detuning $\Delta$. Their effective interaction is a soft-shoulder potential, with a flat repulsion up to a radius $R_C$ related to the highly excited state, and a Van-der-Waals tail at large distances \cite{pupillo2010strongly,Cinti}. Interestingly, quantum cluster phases have been predicted in the high-density and interaction strength regime \cite{henkel_three-dimensional_2010,Cinti,SaccaniMoroniBononsegnisoft}, also in 1D \cite{Mattioli2013,Dalmonte2015}.

In this paper we show the evolution of static and dynamical properties of such 1D soft bosons in the dilute regime. This allows us to test our methodology, as a preliminary study toward higher densities \cite{nostrotonks}. Moreover, the calculation of the dynamical structure factor allows for the uncovering of properties which are hardly ascertainable from static observables.
Following \cite{Cinti}, we consider a shoulder potential, with length $R_C$ and energy $E_C=\hbar^2/m {R_C}^2$ units, where $m$ is the mass, and study the following Hamiltonian in continuous space:
\begin{equation}
 H=-\frac{1}{2}\sum_i \frac{\partial^2}{\partial r^2} + \sum_{i<j}\frac{U}{1 + r^6} \; ,
 \label{eq:shoulderpotential}
\end{equation}
where indexes $i,j$ span all the $N$ particles. In the context of ultracold gases, the interaction strength $U$ can be related to $\Omega$ and $\Delta$ \cite{henkel_three-dimensional_2010}. Eq.~\eqref{eq:shoulderpotential} has to be thought as the effective 1D Hamiltonian which is relevant in an elongated quasi-1D configuration, once the transverse degrees of freedom are reduced to the ground state, due to low temperature and strong trapping (confinement-induced resonance, see \cite{olshanii1998atomic}). We verified that the typical transverse confinement size $a_\perp$ does not affect much the original 3D potential once $a_\perp\lesssim R_C$, which is feasible in current experiments, where both $a_\perp$ and $R_C$ are of order of $10^{2}$ nm \cite{haller_2009}.

\section{Scattering length} \label{sez:tuning}
\begin{figure*}[t]
\begin{minipage}[b]{0.55\textwidth}
    \centering
    \includegraphics[width=\textwidth]{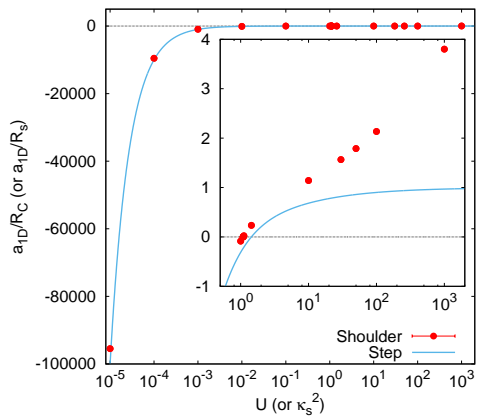}
    \caption{\small Scattering length of the shoulder potential (or step potential with $R_s/R_C=1$). Inset: large strength regime.}\label{fig:should}
\end{minipage}
\begin{minipage}[b]{0.10\textwidth}
    \hspace{0.10\textwidth}
\end{minipage}
\begin{minipage}[b]{0.33\textwidth}
   \centering
    \begin{tabular}[width=\textwidth]{|c|c|}
    \hline
    $U$ & $a_{1D}/R_C$\\
    \hline
    $10^{-5}$ & $ -95470(100)$ \\
    $10^{-4}$ & $ -9550(10)$ \\
    $10^{-3}$ & $ -954.2(9)$ \\
    $10^{-2}$ & $ -94.70(10)$ \\
    $10^{-1}$ & $ -8.743(8)$ \\
    $1.0$ & $ -0.08502(4)$ \\
    $1.0903$ & $ 1.7(4)\cdot10^{-5}$ \\
    $10$ & $ 1.141(1)$ \\
    $30$ & $ 1.565(1)$ \\
    $50$ & $ 1.788(1)$ \\
    $100$ & $ 2.134(2)$ \\
    $1000$ & $ 3.801(3)$ \\
    \hline
    \end{tabular}
    \\\hspace{\textwidth}\\\hspace{\textwidth}\\\hspace{\textwidth}\\\hspace{\textwidth}
    \captionof{table}{\small Shoulder potential scattering length.\\} 
    \label{tab:should}
\end{minipage}
\end{figure*}

Since we consider a dilute, zero temperature system, interaction effects are well described by the scattering length. In 1D, the two-body scattering solution of short-range potentials can be written as $f(r)=\cos\left(k|r|+\delta(k)\right)$ in the region where the potential is sufficiently suppressed. The phase shift has the expansion $\delta(k)=-\frac{\pi}{2}-k a_{\text{1D}}+\mathcal{O}[k^2]$ and the 1D scattering length is defined as $a_{\text{1D}} = -\lim_{k \to 0} \frac{\partial\delta(k)}{\partial k}$. Notice that this definition results in a {\em negative} 1D scattering length for the repulsive Lieb-Liniger model: $V_{LL}(r)=g \delta(r)$, with $g = 2\hbar^2/m|a_{\text{1D}}|$, which is usually parametrized by the dimensionless coupling $\gamma=2/n|a_{\text{1D}}|$ \cite{LiebLinigerI1963}. Moreover, for the hard-rods model, $a_{\text{1D}}>0$ is equal to the hard-core radius \cite{Mazzanti,Girardeau}. For the step potential, defined as $V_s(x)=V_s$ for $x\le R_s$, $V_s(x)=0$ for $x>R_s$, the analytical form of the scattering length is:
\begin{equation}
a_{\text{1D}} = R_s \left( 1 - \frac{1}{\kappa_s\tanh\kappa_s}\right)\;,
\label{eq:a1D}
\end{equation}
where $\kappa_s=\sqrt{m V_s R_s^2/\hbar^2}$. $a_{\text{1D}}$ changes sign at $\kappa_s^2\simeq1.439$ (see Fig. \ref{fig:should}). 

For the shoulder potential $a_{\text{1D}}$ is not known analytically, and we resort to a numerical calculation, which we now describe. We solve the two-body Schr\"odinger equation problem in the $\left[0,l\right]$ interval, 
dividing it into $N_{P}$ subintervals. We use different subinterval lengths: a shorter $dr$ for values of $r$ smaller than $10$ (Region I), 
and a larger interval $dr^\prime$ for the farther region (Region II). This choice allows us to have a better precision in Region I. 
We approximate the kinetic Laplace operator with finite differences and obtain a tridiagonal $(N_{P}\times N_{P})$ matrix, which is asymmetric at the boundary between regions I and II. 
We impose two boundary conditions: $f'(r=0) = 0$, as we want our wave-function $f$ to be symmetric at $r = 0$, and $f(r=l+dr^\prime) = 0$, 
in order to get a node and thus consider the first scattering state. Through a similarity transformation we render the hamiltonian matrix tridiagonal symmetric, 
and then the diagonalization is exploited through LAPACK libraries \cite{laug}. We fit the obtained lowest-energy eigenstate to extract the scattering length for several values of $U$. 
The uncertainty in our procedure is found to be of order $10^{-3}$, by comparing an analogous numerical calculation for the step potential with Eq. \eqref{eq:a1D}. The results are shown in Fig.~\ref{fig:should} and Table~\ref{tab:should}. For the same strength of the potential and $R_s=R_C$, $a_{\text{1D}}$ for the shoulder potential is always greater than for the step potential, 
due to the longer range; however, by increasing $R_s/R_C$ it is possible to obtain a scattering length that is larger than $R_C$. The strength of the shoulder potential corresponding to practically $a_{\text{1D}}\simeq 0$ (Tonks-Girardeau gas \cite{Girardeau,Tonks}) is $U = 1.0903$.

\section{Methods} \label{sez:methodologies}
We use the Shadow Path Integral Ground State (SPIGS) \cite{SPIGS} algorithm, a quantum Monte Carlo method which performs an evolution in imaginary time of a trial state $|\Psi_{T}\rangle$ that has the functional form of the Shadow Wave Function \cite{VitielloSWF}. We simulate from $N=25$ to $N=200$ particles using periodic boundary conditions in a segment of length $L=N/n$, where $n$ is the particle number linear density.
Since we are interested in the low-density regime, we consider the two-body Jastrow form: $\phi(x_1\dots x_N) = \exp\left[-\frac{1}{2}\sum_{i<j} \left(u(x_{ij}) + \chi(x_{ij})\right)\right]$, 
where $x=r$ ($s$) for the real (shadow) particles, and $x_{ij}=|x_i-x_j|$. 
The short-range correlation $\exp\left[-u(x)/2\right]$ is taken to be the solution of the two-body Schr\"{o}dinger equation with an auxiliary step potential, with the boundary condition that the wave-function has zero derivative at a distance $\bar{R}<L/2$. The parameters $V_s, R_s$ of the step potential and $\bar{R}$ are free and need to be optimized. We tune the $V_s$ and $R_s$ so as to have the same 1D scattering length as for the shoulder potential: typically $V_s \simeq U$, $R_s\simeq R_C$, $\bar{R}\simeq 1/n$.
The long-range contribution to the Jastrow factors, $\chi(x)$, which allows for the correct description of phonons, is taken to be of the Reatto-Chester form \cite{ReattoChester}:
\begin{equation}
\chi(x) = - \frac{{\alpha}}{\beta} \log\left[\frac{{\sinh}^{\beta} \left(\frac{\pi}{L k_c} \right) + {\sin}^{\beta}\left(\frac{\pi}{L} x\right)}
{{\sinh}^{\beta} \left(\frac{\pi}{L k_c} \right) + 1}\right] \; ,
\end{equation}
where, in a purely variational calculation, $\alpha$ should be related to the velocity of sound $c$ by $\alpha=\frac{2m c}{\pi\hbar n}$ \cite{ReattoChester}, and the terms involving the cut-off $k_c=2\pi/\bar{R}$ provide a smooth connection to the short-distance behavior, depending on the exponent $\beta$ (typically $\beta$=2, but we found also useful to set $\beta=8$, which depresses more the long-range contribution). A gaussian kernel, $\exp{\left[-\sum_i C{|r_i - s_i|}^2\right]}$, finally couples each real particle to the corresponding shadow one. 

In the large $U$ regime, the described analytic Jastrow does not perform well, requiring a long inefficient imaginary-time projection. We thus resort to a numeric Jastrow, which is the solution of the two-body Schr\"{o}dinger equation in the true shoulder potential. In practice, we use the same algorithm as described in the previous Section, with the only difference that we impose null first derivative at $r=L/2$. We also disable shadow particles in this regime.

\section{Results} \label{sez:main}
In this study, we consider the system described by the Hamiltonian \eqref{eq:shoulderpotential} at low density and for three values of the strength $U$, representative of three well-known models: the weakly interacting Lieb-Liniger (LL) model \cite{LiebLinigerI1963}, for $U=10^{-5}$ and $nR_C=10^{-3}$, corresponding to $a_{\text{1D}}=-95470 R_C$ and $\gamma =0.021$, the Tonks-Girardeau (TG) model \cite{Tonks,Girardeau}, for $U=1.0903$ and $nR_C = 10^{-3}$, corresponding to $a_{\text{1D}}=0$ and $\gamma=\infty$, and the Hard-Rods (HR) model \cite{Mazzanti}, for $U=10^3$ and $nR_C=10^{-2}$, corresponding to $a_{1D} = 3.801$ and $na_{\text{1D}}=0.038$.

For these systems, we calculate the energy per particle $\epsilon=E/(NE_C)$, the pair distribution function $g_2(r)$ and the static structure factor $S(q)$. 
Moreover, we evaluate the imaginary-time intermediate scattering function $F(q,\tau)=\langle \rho(q,\tau)\rho(-q) \rangle$, where $\rho$ is the density fluctuation operator \cite{GIFT}. 
We extract the dynamical structure factor from the integral relation $F(q,\tau)=\int e^{-\omega\tau} S(q,\omega)d\omega$. 
When a parametric model for $S(q,\omega)$, and thus $F(q,\tau)$, is reliable, we solve the previous relation with a fit of $F(q,\tau)$, otherwise we use the Genetic Inversion via Falsification of Theories (GIFT) \cite{GIFT,bertaina_one-dimensional_2016} algorithm, 
which is a well-established statistical analytic continuation approach \cite{bertaina_one-dimensional_2016,rossi_microscopic_2012,nava_dynamic_2013,nava_superfluid_2012,arrigoni_excitation_2013,rota_quantum_2013,molinelli_roton_2016,motta_dynamical_2016}.
 
In the LL case, we obtain the energy $\epsilon=9.834(5)\cdot {10}^{-9}$, compatible with the perturbative expression \cite{Takahashi} $\epsilon \simeq \frac{(nR_C)^2}{2} \left(\gamma - \frac{4}{3\pi}{\gamma}^{3/2} \right)=9.831(1)\cdot {10}^{-9}$, where the uncertainty comes from the errorbar assigned to $a_{\text{1D}}$.
In the TG case, we obtain the energy $\epsilon=1.6442(4) \cdot {10}^{-6}$, consistent with the ideal Fermi gas (IFG) value $\epsilon = (\pi nR_C)^2/6=1.6449 \cdot {10}^{-6}$.
In the HR case we simulate $N=25$ particles and choose a higher density $nR_C = 0.01$, to ease convergence. We obtain the energy $\epsilon=1.7747(2) \cdot {10}^{-4}$, which is fully compatible with the HR result \cite{Girardeau} at $na_{\text{1D}}=0.038$, $\epsilon=\frac{(\pi nR_C)^2}{6(1-na_{\text{1D}})^2}\left(1-\frac{1}{N^2}\right) =1.7747(1) \cdot {10}^{-4}$.

\begin{figure*}[t]
\centering
\begin{subfigure}{0.49\textwidth}
\centering
    \includegraphics[width=\textwidth]{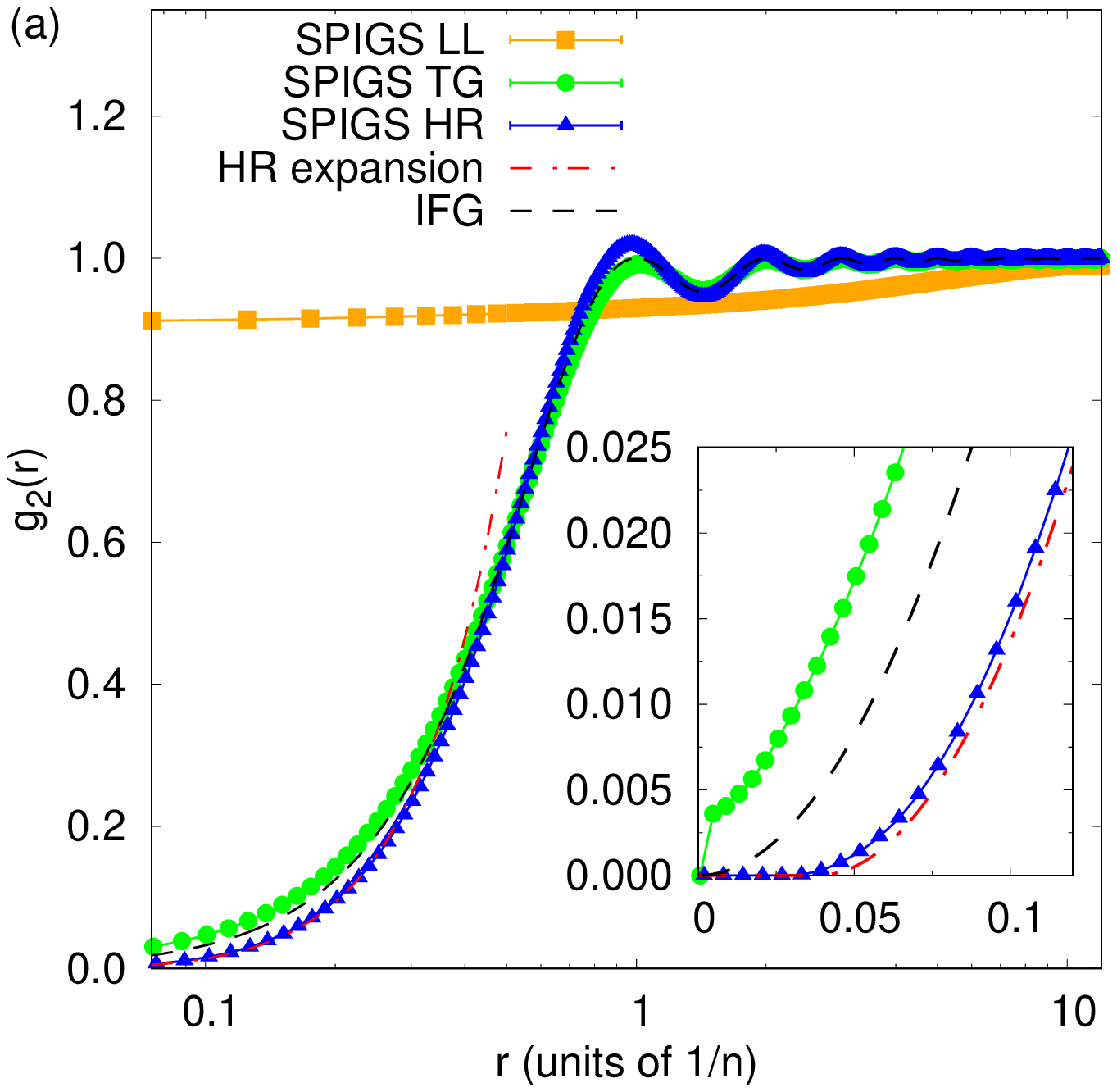}
\end{subfigure}
\begin{subfigure}{0.49\textwidth}
\centering
    \includegraphics[width=\textwidth]{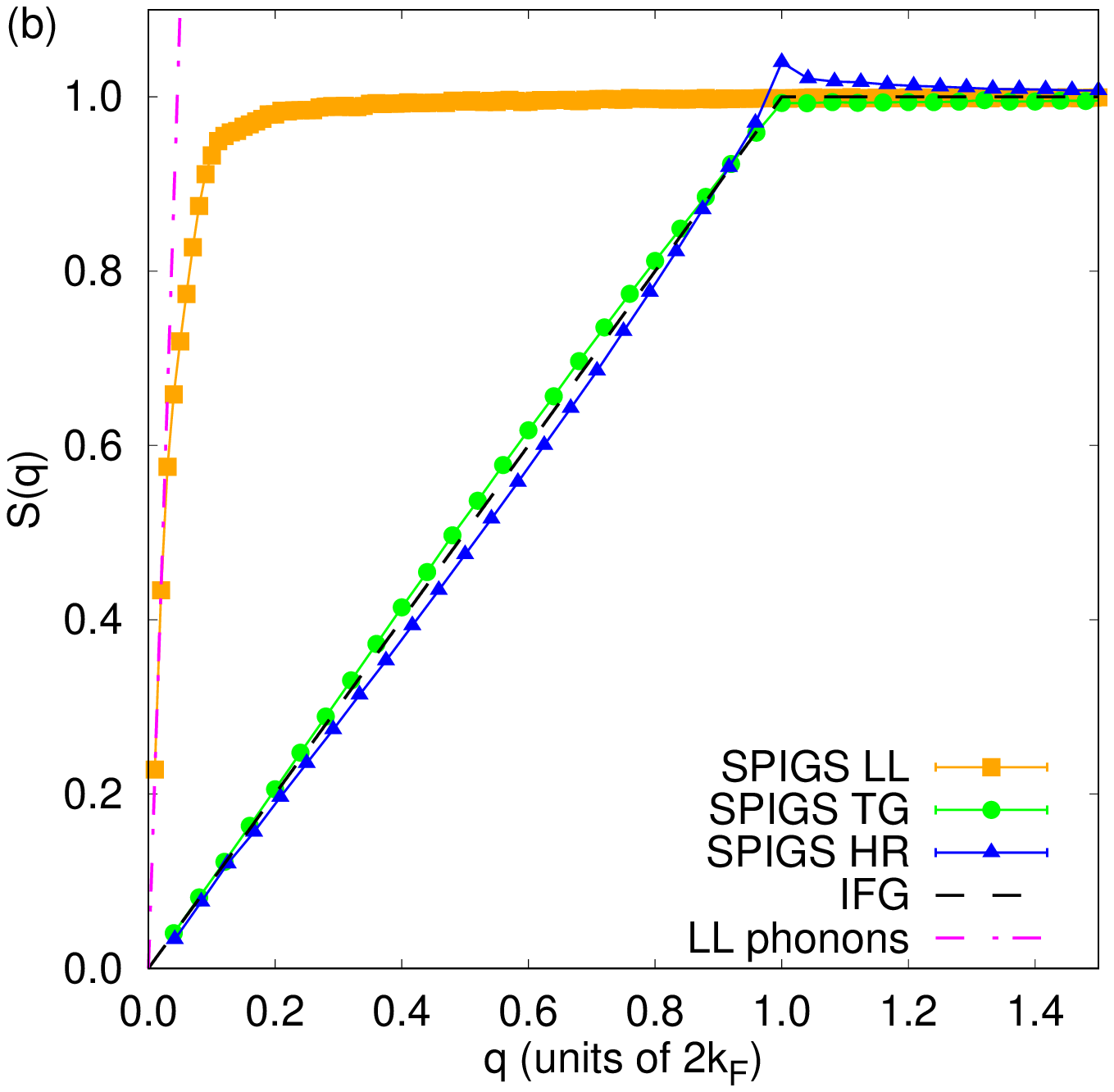}
\end{subfigure}
\caption{\small (a) Pair distribution function and (b) static structure factor in the regimes described in the text. The inset shows the comparison of $g_2$ for the TG and the HR cases to the IFG expression and the low-density HR expansion respectively. (Color figure online)}
\label{fig:PR_gdierreskappax}
\end{figure*}

In Fig.~\ref{fig:PR_gdierreskappax}(a) we show the pair distribution function. One sees the dramatic effect of interaction in going from the weak LL regime to the TG regime, which agrees with the IFG expression $g_2(r)=1-(\sin(k_F x)/k_F x)^2$, where an effective Fermi wave-vector has been defined as $k_F=\pi n$. The HR result is in reasonable agreement with the low- density and distance expansion $g_2(r)\simeq \left[\pi(r-a_{\text{1D}})n/(1-na_{\text{1D}})\right]^2/3$. In the inset we notice that in the TG case at short distances $g_2(r)$ is larger than the value expected from the corresponding model, and drops to zero at $r \simeq R_C$. This is due to the soft-core nature of the shoulder potential. In general, at low density, we expect and indeed observe non-universal effects appearing at very short distances.

In Fig.~\ref{fig:PR_gdierreskappax}(b) we show the static structure factor. For $q\rightarrow0$ the LL case displays a phononic behavior, which is in agreement with the velocity of sound of the weakly interacting LL model $c=\hbar n \gamma^{1/2}/m$. The TG regime compares well with the IFG result $S(q)=q/2k_F$ ($q\le 2k_F$), $S(q)=1$ ($q>2k_F$). In the HR regime, due to the small $n a_{\text{1D}}\simeq0.04$, we observe a small deviation from the IFG result which is evident especially at $q\simeq 2k_F$ \cite{Mazzanti}.

\begin{figure*}[t!]
\centering
\begin{subfigure}{0.49\textwidth}
\centering
    \includegraphics[width=\textwidth]{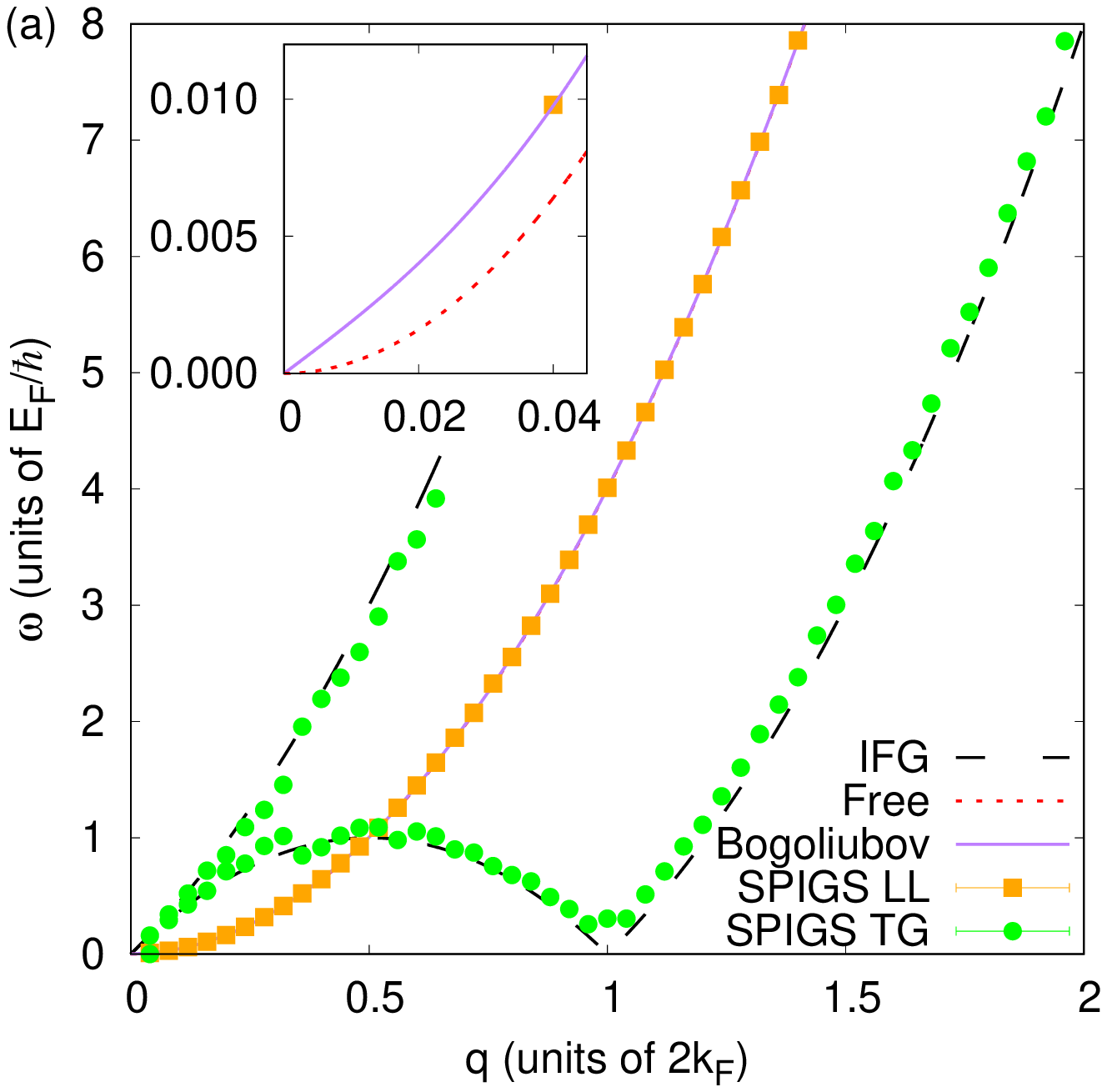}
\end{subfigure}
\begin{subfigure}{0.49\textwidth}
\centering
    \includegraphics[width=\textwidth]{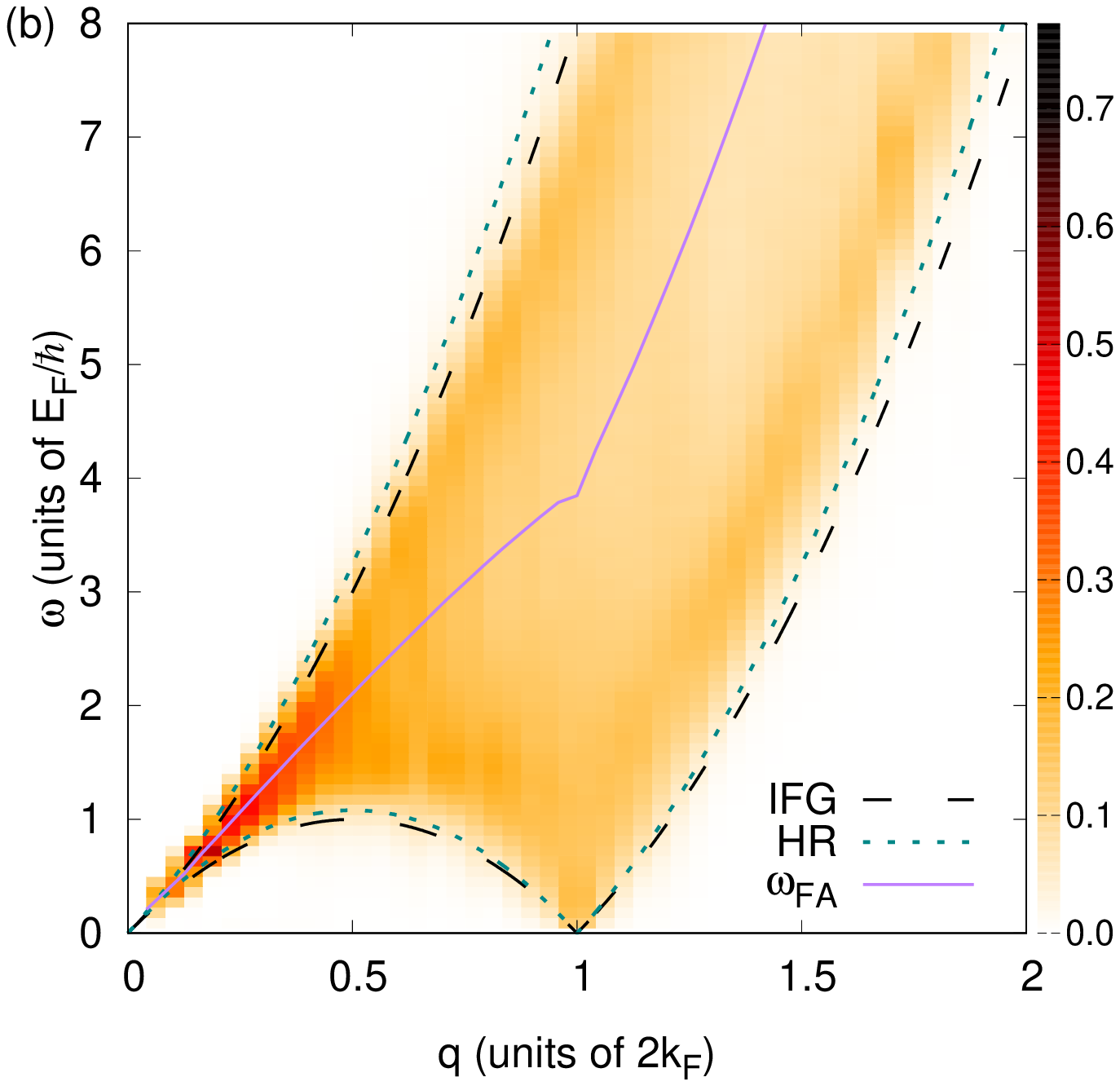}
\end{subfigure}
\caption{\small (a) Dynamical structure factor obtained with simple fits in the LL an TG regimes compared to the Bogoliubov and the free-particle dispersions, and the IFG band. Inset: small-$q$ behavior. (b) Dynamical structure factor in the HR regime obtained with GIFT, in units of $1/E_F$, compared to Feynman's approximation and the analytical particle-hole boundaries of the IFG and HR \cite{bertaina_one-dimensional_2016,motta_dynamical_2016} models. (Color figure online)} \label{fig:PR_spectra}
\end{figure*}

In Fig.~\ref{fig:PR_spectra} we show the dynamical structure factor. Units of momentum $2k_F$ and frequency $E_F/\hbar=\hbar k_F^2/2m$ are most useful to compare to IFG predictions. Since the LL case is very weakly interacting, a single decaying exponential is sufficient to fit the calculated $F(q,\tau)=S(q)e^{-\omega_0\tau}$. In panel (a) one sees that the resulting spectrum is almost free-particle-like, but in the inset the linear long-wavelength Bogoliubov dispersion is shown to be consistent with our data. In the TG case, $S(q,\omega)$ should be constant within the particle-hole band delimited by the boundaries $\omega_{\text{IFG}}^\pm(q)=\left|\frac{q k_F}{m}\pm\frac{\hbar q^2}{2m}\right|$. We indeed easily fit $F(q,\tau)=S(q)(e^{-\omega_1\tau}-e^{-\omega_2\tau})/[\tau(\omega_2-\omega_1)]$, identifying two boundary frequencies $\omega_{1,2}$ that compare well with the IFG result. We ascribe the discrepancy at some momenta, for example at $q=2k_F$, to finite size effects in the lower frequency threshold, which is known to scale at this momentum as $1/N$. The complete evolution of $S(q,\omega)$ in the LL model can be found in \cite{CauxCalabrese}.
In the HR regime we use the GIFT algorithm to estimate $S(q,\omega)$, since no simple form of the spectrum is expected \cite{motta_dynamical_2016}.
In panel (b), the resulting color plot is shown, compared to Feynman's approximation $\omega_{\text{FA}}(q)=\hbar q^2/[2mS(q)]$ \cite{feynman_1954}, the IFG particle-hole band and the renormalized HR particle-hole band $\omega_{\text{HR}}^\pm(q)=\omega_{\text{IFG}}^\pm(q)/(1-na_\text{1D})^2$, which can be derived from nonlinear Luttinger theory \cite{bertaina_one-dimensional_2016,motta_dynamical_2016,imambekov_2009}. The support of the spectrum is clearly identified by the GIFT algorithm, and, at this level of precision, the spectrum is not peaked along Feynman's approximation, but along the lower HR threshold $\omega_{\text{HR}}^-(q)$, as one expects for a system which is less compressible than the IFG \cite{Mazzanti,bertaina_one-dimensional_2016}.

\section{Conclusions} \label{sez:conclusions}
We have presented the Quantum Monte Carlo study of a system of dilute one-dimensional soft spheres at zero temperature. We have investigated both the statics and the dynamics of the system. A remarkable part of this work has been testing the SPIGS method with a new potential and wave-function, by using well-known one-dimensional models as a check of our program in the dilute regime. In order to make the comparison with such models, we performed the calculation of the shoulder potential scattering length for varying strength $U$. We managed to recover static and dynamical properties of the the Lieb-Liniger model, both in the weakly interacting regime, and in the strongly interacting Tonks-Girardeau limit, where the behavior is similar to the Ideal Fermi Gas. In the strongly interacting regime, the system is well described by the Hard-Rods model.
This study allows us to consider higher-density regimes, for example fixing $a_\text{1D}=0$ \cite{nostrotonks} or increasing $U$ to describe cluster phases \cite{Mattioli2013}.

\begin{acknowledgements}
We acknowledge the CINECA ISCRA award SoftDyn, for the availability of high-performance computing resources and support, and D.E. Pini for useful discussions. G.B. acknowledges support from the University of Milan through Grant No. 620, Linea 2 (2015).
\end{acknowledgements}


\begin{thebibliography}{10}
\providecommand{\url}[1]{{#1}}
\providecommand{\urlprefix}{URL }
\expandafter\ifx\csname urlstyle\endcsname\relax
  \providecommand{\doi}[1]{DOI \discretionary{}{}{}#1}\else
  \providecommand{\doi}{DOI \discretionary{}{}{}\begingroup
  \urlstyle{rm}\Url}\fi

\bibitem{cazalilla_one_2011}
M.~Cazalilla, et~al., Rev. Mod. Phys. \textbf{83}(4), 1405 (2011)

\bibitem{schaus_observation_2012}
P.~{Schau\ss{}}, et~al., Nature \textbf{491}(7422), 87 (2012)

\bibitem{henkel_three-dimensional_2010}
N.~Henkel, R.~Nath, T.~Pohl, Phys. Rev. Lett. \textbf{104}(19), 195302 (2010)

\bibitem{Cinti}
F.~Cinti, T.~Macrì, W.~Lechner, G.~Pupillo, T.~Pohl, {Nat}. {Commun}.
  \textbf{5}, 3235 (2014)

\bibitem{SaccaniMoroniBononsegnisoft}
S.~Saccani, S.~Moroni, M.~Boninsegni, {Phys.} {Rev.} {Lett.} \textbf{108}(17),
  175301 (2012)

\bibitem{zeiher_many-body_2016}
J.~Zeiher et al., Nat. Phys. (2016) \doi{http://dx.doi.org/10.1038/nphys3835}.


\bibitem{pupillo2010strongly}
G.~Pupillo, et~al., {Phys.} {Rev.} {Lett.} \textbf{104}, 223002 (2010)

\bibitem{Mattioli2013}
M.~Mattioli, M.~Dalmonte, W.~Lechner, G.~Pupillo, Phys. {Rev.} {Lett.}
  \textbf{111}, 165302 (2013)

\bibitem{Dalmonte2015}
M. Dalmonte, W. Lechner, Z. Cai, M. Mattioli, A.M. L\"auchli, and G. Pupillo, {Phys. Rev. B} \textbf{92}, 045106 (2015).

\bibitem{nostrotonks}
M.~Teruzzi, D.~Galli, D.~Pini, G.~Bertaina,  {\em in preparation}

\bibitem{olshanii1998atomic}
M.~Olshanii, {Phys.} {Rev.} {Lett.} \textbf{81}(5), 938 (1998)

\bibitem{haller_2009}
E.~Haller, et al., Science \textbf{325}, 1224 (2009)

\bibitem{LiebLinigerI1963}
E.H. Lieb, W.~Liniger, {Phys.} {Rev.} \textbf{130}(4), 1605 (1963)

\bibitem{Mazzanti}
F.~Mazzanti, G.E. Astrakharchik, J.~Boronat, J.~Casulleras, {Phys}. {Rev}. {Lett}. \textbf{100}, 020401 (2008)

\bibitem{Girardeau}
M.~Girardeau, {J.} {Math.} {Phys.} \textbf{1}(6), 516 (1960)

\bibitem{laug}
E.~Anderson, et~al., \emph{{LAPACK} Users' Guide} (SIAM, Philadelphia, PA, 1999)

\bibitem{Tonks}
L.~Tonks, {Phys}. {Rev}. \textbf{50}, 955 (1936)

\bibitem{SPIGS}
D.E. Galli, L.~Reatto, {Mol.} {Phys.} \textbf{101}(11), 1697 (2003)

\bibitem{VitielloSWF}
S.~Vitiello, K.~Runge, M.H. Kalos, {Phys}. {Rev}. {Lett}. \textbf{60}, 1970 (1988)

\bibitem{ReattoChester}
L.~Reatto, G.V. Chester, {Phys.} {Rev.} \textbf{155}, 88 (1967)

\bibitem{GIFT}
E.~Vitali, M.~Rossi, L.~Reatto, D.E. Galli, {Phys.} {Rev.} B \textbf{82}, 174510 (2010)

\bibitem{bertaina_one-dimensional_2016}
G.~Bertaina, M.~Motta, M.~Rossi, E.~Vitali, D.~Galli, Phys. Rev. Lett. \textbf{116}, 135302 (2016)

\bibitem{rossi_microscopic_2012}
M.~Rossi, E.~Vitali, L.~Reatto, D.~Galli, Phys. Rev. B \textbf{85}, 014525 (2012)

\bibitem{nava_dynamic_2013}
M.~Nava, D.~Galli, S.~Moroni, E.~Vitali, Phys. Rev. B \textbf{87}, 144506 (2013)

\bibitem{nava_superfluid_2012}
M.~Nava, D.~Galli, M.~Cole, L.~Reatto, J. Low Temp. Phys. \textbf{171}(5-6), 699 (2012)

\bibitem{arrigoni_excitation_2013}
F.~Arrigoni, E.~Vitali, D.E.~Galli, and L.~Reatto, Low Temp. Phys. \textbf{39}(9), 793 (2013). [Fiz. Nizk. Temp. \textbf{39}, 1021-1030 (2013)]

\bibitem{rota_quantum_2013}
R.~Rota, F.~Tramonto, D.~Galli, S.~Giorgini, Phys. Rev. B \textbf{88}, 214505 (2013)

\bibitem{molinelli_roton_2016}
S.~Molinelli, D.~Galli, L.~Reatto, M.~Motta, J. Low Temp. Phys. \textbf{185}, 39 (2016)

\bibitem{motta_dynamical_2016}
M.~Motta, E.~Vitali, M.~Rossi, D.E.~Galli, and G.~Bertaina, Phys. Rev. A \textbf{94}, 043627 (2016)

\bibitem{Takahashi}
M.~Takahashi, {Prog.} {Theor.} {Phys.} \textbf{53}(2), 386 (1975)

\bibitem{CauxCalabrese}
J.S. Caux, P.~Calabrese, {Phys.} {Rev.} {A} \textbf{74}, 031605 (2006)

\bibitem{feynman_1954}
R.P.~Feynman, Phys. Rev. \textbf{94}, 262 (1954)

\bibitem{imambekov_2009}
A.~Imambekov and L.I.~Glazman, Science \textbf{323}, 228 (2009)


\end{thebibliography}
\end{document}